\documentclass{aa}
\usepackage[dvips]{graphicx} 
\usepackage[dvips]{epsfig}
\usepackage{natbib}
\usepackage{multirow}
\bibpunct{(}{)}{;}{a}{}{,} 
\newcommand{\Teff}{$T_\mathrm{eff}$}
\newcommand{\PEGASEHR}{P\'EGASE-HR}
\newcommand{\Zsol}{Z$_\odot$}

\begin{document}
\title{Evolutionary synthesis of galaxies at high spectral\\resolution with the code \PEGASEHR}
\subtitle{Metallicity and age tracers}

\authorrunning{Le Borgne et al.} 
\titlerunning{Galaxy evolution at high spectral resolution} 

\author{D. Le Borgne\inst{1,2}, B. Rocca-Volmerange\inst{1,7}, 
P. Prugniel \inst{3}, A. Lan\c con \inst{4}, M. Fioc\inst{1,6}, C. Soubiran \inst{5}}

\offprints{Damien Le Borgne, \email{leborgne@iap.fr}}

\institute{Institut d'Astrophysique de Paris, 98\,{\it bis}, Boulevard Arago, F-75014 Paris, France 
\and Dept. of Astronomy \& Astrophysics, University of Toronto, 
60 St. George Street, Toronto, ON M5S 3H8, Canada
\and CRAL-Observatoire de Lyon,  9 av. C. Andr\'e, F-69561 Saint-Genis Laval, France
\and Observatoire de Strasbourg,  11 rue de l'Universit\'e, 67000 Strasbourg, France
\and Observatoire de Bordeaux,  BP 89, 33270 Floirac, France
\and Universit\'e Pierre et Marie Curie, 4 place Jussieu, 75005 Paris, France
\and Universit\'e Paris-Sud, 91405 Orsay, France}
\date{Received 2 May 2003 / Accepted 23 June 2004}

\abstract{We present \PEGASEHR, a new stellar population synthesis
  program generating high resolution spectra ($R=10\,000$) over the
  optical range $\lambda\lambda$ = 400--680~nm. It links the
  spectro-photometric model of galaxy evolution P\'EGASE.2
  \citep{FRV97} to an updated version of the \'ELODIE library of
  stellar spectra observed with the 193~cm telescope at the
  Observatoire de Haute-Provence \citep{ELODIE}.  The \'ELODIE star
  set gives a fairly complete coverage of the Hertzprung-Russell (HR) diagram and
  makes it possible to synthesize populations in the range $\textrm{[Fe/H]}=-2$ to $+0.4$.
  This code is an exceptional tool for exploring signatures of
  metallicity, age, and kinematics.  We focus on a detailed study of
  the sensitivity to age and metallicity of the high-resolution
  stellar absorption lines and of the classical metallic indices
  proposed until now to solve the age-metallicity
  degeneracy.  Validity tests on several stellar lines are performed
  by comparing our predictions for Lick indices to the models of other
  groups.  The comparison with the lower resolution library BaSeL
  \citep{Lejeune97} confirms the quality of the \'ELODIE library when
  used for single stellar populations (SSPs) from $10^7$ to
  $2\times10^{10}$~yr.  Predictions for the evolved populations of
  globular clusters and elliptical galaxies are given and compared to
  observational data.  Two new high-resolution indices are proposed
  around the $H\gamma$ line. They should prove useful in the analysis
  of spectra from the new generation of telescopes and spectrographs.
   } \maketitle
\section{Introduction}
The rapidly increasing number of high-quality spectroscopic galaxy
surveys makes it necessary to improve the galaxy evolution models
used for their interpretation.
To derive the history of the star formation and the chemical evolution of
a stellar population from its line-of-sight integrated spectrum one can either 
fit spectrophotometric indices and colors with a model, or fit
directly the observed
spectrum with a synthetic spectral energy distribution (SED).

Spectrophotometric indices characterize the strengths of spectral features
that are sensitive to the age or to the metallicity of  a stellar
population, and generally to both.  The indices of \citet{Rose1994}
and the Lick indices \citep{Worthey1994,KD1998,Trager} are the most
widely used. Various models allow us to predict the evolution of these
indices. They are based on assumptions for the stellar evolution
(evolutionary tracks) and for the history of stellar formation
(initial mass function, star formation rate\ldots). They usually
require preliminary measurements of the indices in a library of stars.
The most recent models \citep[e.g.][]{Thomasetal2002} take into account the
non-solar abundances resulting from the different processes of metal
enrichment by using the  response functions of \citet{TripiccoBell1995}.
Other families of models compute the SED, thus allowing the user to
measure indices {\it a posteriori} if the spectral resolution is
sufficient \citep[e.g.][]{FRV97,Leitherer1999,Eisenstein2003,BC03}.
The P\'EGASE code is one of the latter, and its latest
version P\'EGASE.2 \citep{PEGASE2} is the basis for the present study.

One of the major issues in studying stellar populations is the
age-metallicity degeneracy \citep{Worthey1994}.  Most
indices are sensitive to both metallicity and age: a
younger age may be confused with a lower metallicity. 
The comparison of indices featuring Balmer lines with indices of
metallic lines may in principle break the degeneracy. But in practice
the tests are difficult and give ambiguous results, partly because of
the contamination by nebular emission, in particular for H$\alpha$ and
H$\beta$ \citep{Gonzalez1993,Wortheyetal1997,Kuntschner2001}.

The degeneracy can also be partly lifted either by using broad-band colors
and indices in an extended range of wavelengths
\citep{G93,Wortheyetal1994} or, more efficiently, by defining indices at a higher
spectral resolution \citep{Jones1995}.  P\'EGASE.2, which is based on
the library of \citet{Lejeune97,Lejeune98}, is well suited to the first
approach. Until recently, such SED-predicting algorithms were
unable to reach high spectral resolutions because they lacked adequate
stellar spectral libraries.

In a pioneering work, \citet{Vazdekis1999} demonstrated the 
potentiality of using SEDs at higher resolution. They used the
library of \citet{Jones1998}, thus reaching $R=2000\textrm{--}3000$
(the Lick resolution is about 500), and were able to define narrower 
age-sensitive indices around H$\gamma$ (with band-widths of 10--20~\AA).

At high resolution, the analysis of a galaxy spectrum must take into
account another phenomenon: the internal kinematics broadens the lines
and reduces the apparent resolution of the spectrum. Traditionally,
measurements of spectrophotometric indices are corrected for velocity
dispersion \citep[see e.g.][]{Golevetal1999}, but for giant elliptical
galaxies with typical velocity dispersions $\sigma$ close to  
300~km\,s$^{-1}$, the correction would introduce unacceptable errors for
narrow indices. For dwarf galaxies ($\sigma < 60$~km\,s$^{-1}$) and
globular clusters ($\sigma < 10$~km\,s$^{-1}$), corrections would
remain acceptable for indices defined on 2~\AA-wide passbands.

Synthetic high resolution SEDs are potentially very useful 
for probing the internal kinematics. They may indeed replace the traditional 
stellar templates and provide both a measurement of the line-of-sight
velocity distribution (LOSVD) and a constraint on the stellar
population. Recently, the
\citet{Vazdekis1999} model has been coupled to the \ion{Ca}{ii} library of
\citet{Cenarroetal2001} to determine the LOSVD profiles of bulges of
spiral galaxies \citep{FalconBarrosoetal2003}. The LOSVDs are
determined by deconvolving the observed
spectrum with a template spectrum resulting from the population
synthesis program.
Using synthesized spectra to study the internal kinematics is not
straightforward. Indeed, it is first necessary to convolve the template
with the spectral instrumental response, which is generally neither Gaussian nor
constant over the whole wavelength range. To do this, the template
spectrum must have a significantly higher resolution than the
observed one.

To address the questions above, we have coupled the last version of P\'EGASE
to a library of high resolution ($R=10\,000$) stellar spectra. 
In this paper we present the resulting code, \PEGASEHR, and focus on its predictions 
of line indices that are sensitive to age and metallicity.

In Sect.~2, we recall the main characteristics of P\'EGASE.2. We also
briefly describe the main features of the updated stellar library \'ELODIE
\citep{ELODIE} as well as the ``interpolator'', the method used to convert the
set of \'ELODIE stellar spectra to a grid of spectra with
the regularly spaced  stellar parameters of the BaSeL library.  The
synthetic high-resolution spectra of SSPs are then
compared to the corresponding low-resolution spectra of P\'EGASE.2. 
The variations of the high resolution spectra around commonly
exploited lines are illustrated, as a function of both age and metallicity. 
In Sect.~3, the integrated fluxes, colors and line indices predicted with
P\'EGASE-HR are validated by comparison with previous works.
In Sect.~4, as a further test of the model predictions, we compare them
to observations of globular clusters and elliptical galaxies. 
A systematic search for new narrow indices is
described in Sect.~5. Perspectives for the new generation of
instruments are suggested in the conclusion.

\section{\PEGASEHR: the coupling of P\'EGASE.2  and \'ELODIE}
\subsection{P\'EGASE.2, a spectral evolution code}
The P\'EGASE.2 code\footnote{ Version~2 of P\'EGASE, \emph{Projet d'\'Etude des GAlaxies
    par Synth\`ese \'Evolutive} in French. The code is available at
  \texttt{http://www.iap.fr/pegase/}.}
is aimed at modeling the spectral  evolution of galaxies.  It is based
on stellar evolutionary tracks from the ``Padova'' group, extended to
the thermally pulsating asymptotic giant branch (AGB) and post-AGB
phases; these tracks cover all the masses, metallicities and phases of
interest for galaxy spectral synthesis.  For a given evolutionary
scenario (typically characterized by a star formation law, an initial
mass function and, possibly, infall or galactic winds), the code
consistently computes the star formation rate and the metallicity of
gas and stars at any time.  The nebular component (continuum and
lines) due to \ion{H}{ii} regions is roughly calculated 
and added to the stellar component.  Depending on the type of galaxy
(disk or spheroidal), the attenuation of the spectrum by dust is then
computed using the output of a radiative transfer code; this code
takes into account scattering \citep{These_Michel}.

P\'EGASE.2 uses the BaSeL \citep{Lejeune97,Lejeune98} library of stellar
spectra and can therefore synthesize low-resolution ($R\simeq200$)
ultraviolet to near-infrared spectra of Hubble sequence galaxies as
well as of starbursts. In \PEGASEHR, the BaSeL library is replaced  
by a grid of spectra interpolated from the  high-resolution \'ELODIE
library of stellar spectra.

\subsection{\'ELODIE, a high resolution library of stellar spectra}
The \'ELODIE library is a stellar database of 1959 spectra for
1503 stars, observed with the \'echelle spectrograph \'ELODIE on the
193~cm telescope at the Observatoire de Haute Provence. Previous
versions of the library were presented in \citet{ELODIE_first} and
\citet{ELODIE,ELODIE_cat}.  It has been updated for the present work by
doubling the number of spectra, which greatly improved the coverage of the
parameter space (in effective temperature, surface gravity, and metallicity).
The data reduction has also been improved, in particular the flux
calibration, and the wavelength range has been extended to
$\lambda\lambda = 400\textrm{--}680$~nm.  For the purpose of population
synthesis, the original resolution ($R=42\,000$) has been reduced to
$R=10\,000$ at $\lambda=550$~nm, or more precisely to a gaussian
instrumental profile of FWHM$\simeq0.55$~\AA{} over the whole range of
wavelengths.  The typical signal-to-noise (S/N) ratio of the spectra is 500/\AA.  
The connection of several \'echelle spectra, each one
 defined in a small wavelength range, is needed 
to build a single spectrum over the entire range $\lambda\lambda = 400\textrm{--}680$~nm. 
The complex procedure used to perform this task leads to an overall (broad-band) photometric
precision of 2\%. The precision
reaches 0.5\% when the \'ELODIE spectra are normalized to a
low resolution continuum, as
described in detail in \citet{ELODIE}.

The atmospheric parameters of the stars (\Teff, $g$, [Fe/H]) are taken
from the literature in an up-to-date version of \citet{Cayrel}, when
available.  Quality weights are applied to the parameters when
several estimations were available \citep[for details,
see][]{ELODIE}\footnote{and also
http://www.obs.u-bordeaux1.fr/public/astro/
CSO/elodie\_library.html}.  Otherwise they are estimated with the
TGMET procedure, which consists in least square fits of the target
spectrum to reference spectra with known atmospheric parameters
\citep{TGMET,TGMET2}.  Unlike
\citet{Wortheyetal1994} we chose  not to calibrate the temperature of  the giant stars on their spectral
type as given by \citet{Ridgwayetal1980} because this study is contested in several
recent papers \citep{Sekiguchietal2000,Houdasheltetal2000}.  As a
consequence, our estimated \Teff{} for the giant stars is globally
smaller by about 5\% than the values given by
\citet{Wortheyetal1994}. This will be discussed in
Sect.~\ref{section:Lick}.

\begin{figure}[!tbf]
  \centering
  \includegraphics{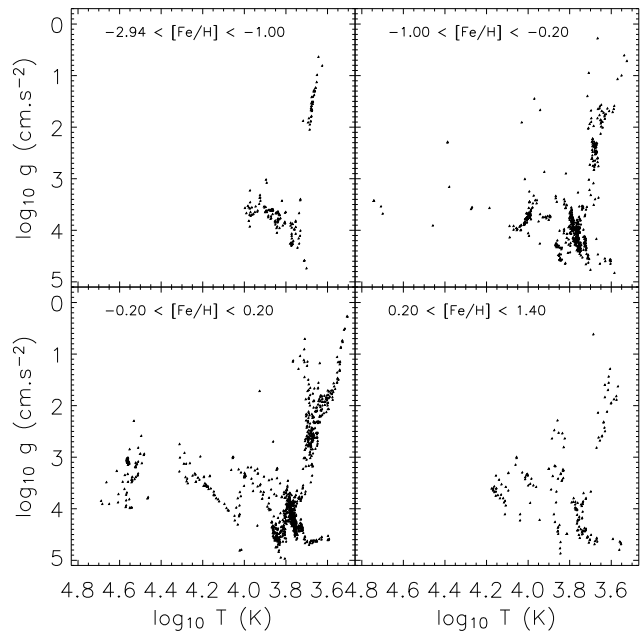}
  \caption{Distribution in the \Teff - $\log_\mathrm{10} g$ 
    diagram of the 1503 stars in the \'ELODIE library in four bins of metallicity.}
  \label{figure:HRdiag}
\end{figure}

The HR diagram coverage with the estimated parameters is extensive:
$0.27<$~$\log_\mathrm{10} (g\ \mathrm{[cm.s^{-2}]})$~$<4.97$ and
$3\,185$~K$ \le $\Teff$\le 55\,200$~K, with $-3.21 \le $[Fe/H]$ \le
1.62$. This coverage is illustrated in Fig.~\ref{figure:HRdiag}.

The up-to-date version of the library is
available
online\footnote{http://www-obs.univ-lyon1.fr/\~{}prugniel/soubiran/}.
Fully reduced spectra, as well as the estimated stellar parameters, are provided.

\begin{center}
  \begin{figure*}[!tbf]
    \begin{center}
      \includegraphics{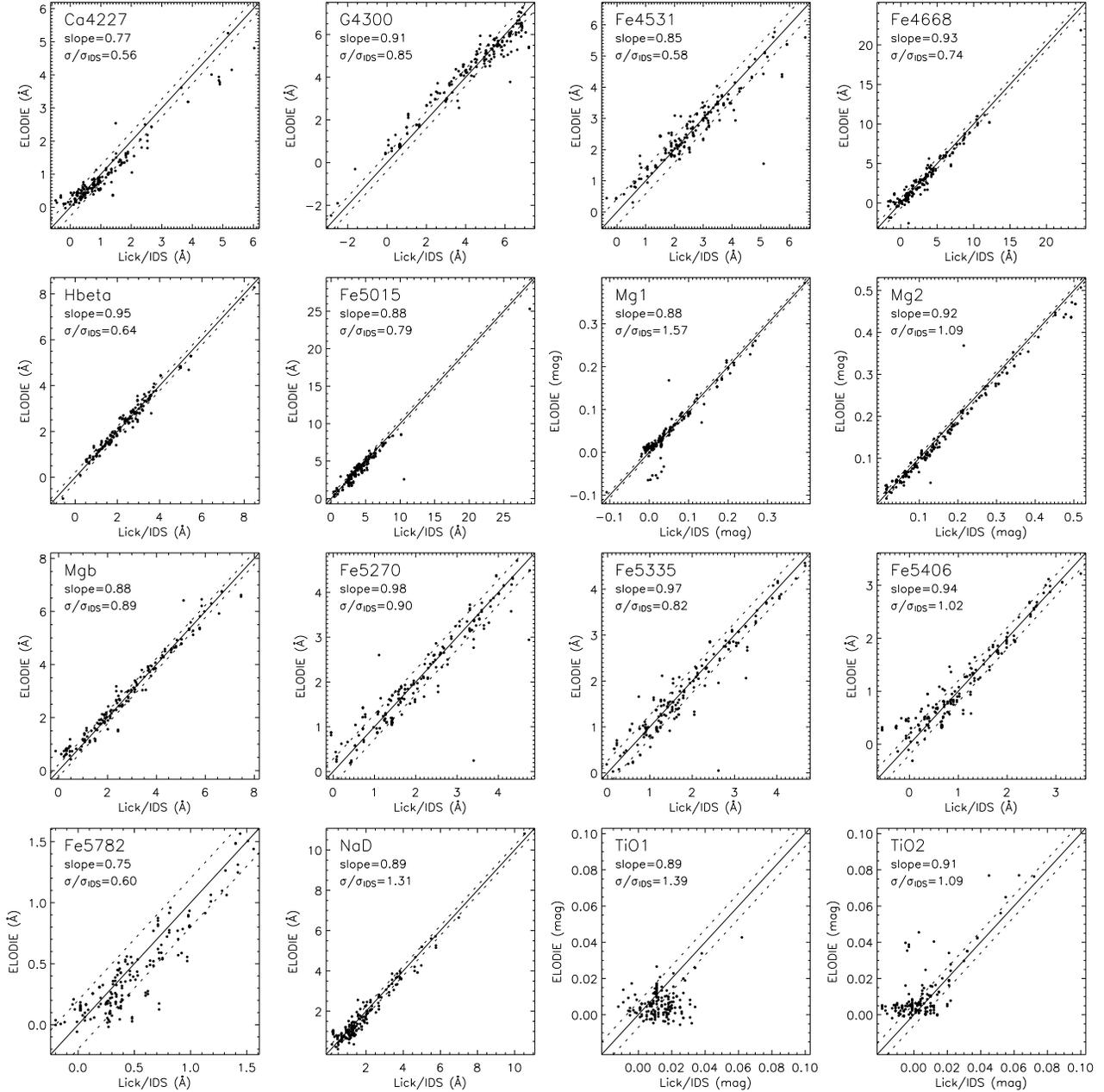}
    \end{center}
    \caption{Comparison of Lick indices measured on the \'ELODIE
    spectra with the reference Lick/IDS indices
    \citep{Wortheyetal1994} for 187 stars in common. The dotted lines
    are $y=x\pm \sigma_{\mathrm{IDS}}$ where $\sigma_\mathrm{IDS}$ is the uncertainty on
    the Lick/IDS indices. For each panel, we give the
    slope of the linear regression and  the dispersion $\sigma$ around it.
    Ideally, $\sigma / \sigma_\mathrm{IDS} \simeq 1.$}
    \label{figure:lick}
  \end{figure*}
\end{center}

Since the number of spectra included in the \'ELODIE library has
doubled since \citet{ELODIE}, we update here the comparison of the
Lick indices measured on the stellar spectra with the reference
indices from \citet{Wortheyetal1994} for the 187 stars in
common. Figure~\ref{figure:lick} illustrates this comparison for 16
indices, showing the generally good agreement between the two sets of
spectra. However, a closer look reveals a general trend for our
strongest indices to be weaker than the reference indices, which is
expressed by the slopes being smaller than unity.
 To investigate this bias, we also compared
our spectra to the Jones library \citep{Leitherer96,Jones1998} for the 314 stars in
common. This library has the advantage of being well calibrated in
wavelength, as opposed to the Lick/IDS spectra of \citet{Wortheyetal1994}. When
compared at the same resolution (1.8~\AA), the two sets of spectra agree
remarkably well. 
\begin{table}[!tbf]
\begin{center}
\caption[]{Comparison of Lick indices measured on \'ELODIE stellar spectra with indices 
    from other databases of stellar spectra, for stars in common. 
    The indices for \citet{Wortheyetal1994} were taken from the table published in their paper. 
    The indices for the Jones library were measured by ourselves (see text).
    The values are the mean offset (\'ELODIE$-$other~database), the slope 
    of the linear regression and  the dispersion $\sigma$ around it (see Fig.~\ref{figure:lick}).}
\begin{center}
\begin{tabular}{ l  rcc c@{} rcc}
\hline
\hline
    \multirow{2}*{Index}&\multicolumn{3}{c}{Worthey et al. 1994} && \multicolumn{3}{c}{Jones library} \\
\cline{2-4}\cline{6-8}
                        &   offset&    slope& $\sigma$      &&   offset&    slope& $\sigma$    \\
\hline
    Ca4227              &  0.038  &  0.772  &  0.151       &&   0.023  &  0.890  &  0.074\\
     G4300              &  0.025  &  0.910  &  0.333       &&   0.094  &  0.942  &  0.191\\
    Fe4531              &  0.017  &  0.849  &  0.245       &&   &  &\\
    Fe4668              &  0.037  &  0.931  &  0.476       &&   &  &\\
     H$\beta$           &  0.022  &  0.947  &  0.140       &&   0.002  &  0.958  &  0.069\\
    Fe5015              & -0.136  &  0.884  &  0.358       &&  -0.001  &  0.966  &  0.163\\
       Mg1              & -0.001  &  0.878  &  0.011       &&  -0.003  &  0.981  &  0.011\\
       Mg2              &  0.001  &  0.918  &  0.009       &&   0.000  &  0.963  &  0.007\\
       Mg$_b$           & -0.037  &  0.877  &  0.204       &&   0.019  &  0.996  &  0.070\\
    Fe5270              & -0.033  &  0.978  &  0.252       &&   0.012  &  0.991  &  0.057\\
    Fe5335              & -0.017  &  0.974  &  0.214       &&   0.003  &  0.947  &  0.061\\
    Fe5406              & -0.003  &  0.937  &  0.203       &&   0.008  &  0.994  &  0.047\\
    Fe5782              & -0.011  &  0.745  &  0.120       &&   &  &\\
       NaD              &  0.073  &  0.895  &  0.314       &&   &  &\\
      TiO1              & -0.001  &  0.889  &  0.010       &&   &  &\\
      TiO2              &  0.001  &  0.909  &  0.007       &&   &  &\\                     
\hline
\end{tabular}
\end{center}
\label{table:wj}
\end{center}
\end{table}

Table~\ref{table:wj} gives the quantitative comparison of these three
libraries in terms of Lick indices. The indices of the Jones library
were measured by ourselves when computing this table. The good
agreement between \'ELODIE and Jones spectra is shown by the slopes
being closer to unity and the smaller dispersions $\sigma$. It is worth
noticing that the indices published by Jones differ from the ones we
measure from their spectra: they seem to have been artificially corrected in
order to match the \citet{Wortheyetal1994} indices by changing the
overall slope.  Our routine used to measure Lick indices was tested
successfully on seven stars provided by G. Worthey. To measure the
indices on the higher resolution spectra of the \'ELODIE or the Jones
libraries, we use the wavelength-dependent resolution given in
\citet{Wortheyetal1997}.

The small bias observed between the measurements made
on our library and the Jones library is probably due to the imperfect
calibration of the spectra in either library.  The reason for the
discrepancy between our measurements and the Lick/IDS ones is still
unclear, but it might be attributed to a systematic effect in the
wavelength scale of the Lick library.

In the following, we do not
correct any of our indices for this very small bias. 
Section~\ref{section:Lick} will show that its consequences 
on SSPs are usually small, although they may become important
when red giants dominate the spectra.

To use the set of \'ELODIE spectra as an input stellar library to
P\'EGASE, the spectra were interpolated on the BaSeL
\citep{Lejeune97} grid of parameters (\Teff, $\log_\mathrm{10} g$,
[Fe/H]) with piecewise polynomials  similar to the Lick fitting
functions \citep{Worthey1994}, in various regions of the HR diagram.
The interpolated spectra were then normalized to match the mean BaSeL
  fluxes in the band 5500--6000~\AA{}. The detailed description of the
interpolator, as well as the comparison of the BaSeL library with 
the interpolated stellar library, are described on a web 
site\footnote{http://www.obs.u-bordeaux1.fr/public/astro/CSO/ elodie\_library.html}
and will be published in a forthcoming paper \citep{interpolateur}.

The range of stellar parameters in the BaSeL library being much
larger than the range of parameters covered by the \'ELODIE library, we
chose not to use the extrapolated spectra of the interpolated
library. To exclude the extrapolated spectra, we
set a minimum level for the density of real stellar spectra at each
point of the BaSeL grid. The resulting interpolated library then 
contains fewer spectra than the BaSeL library (2690 instead of 4422), 
but each of its spectra is reliable. A minor modification in the P\' EGASE 
algorithm was necessary to use this new grid.
%
%
\subsection{SSPs at high spectral resolution}
\begin{center}
  \begin{figure*}[!tbf]
    \begin{center}
      \includegraphics{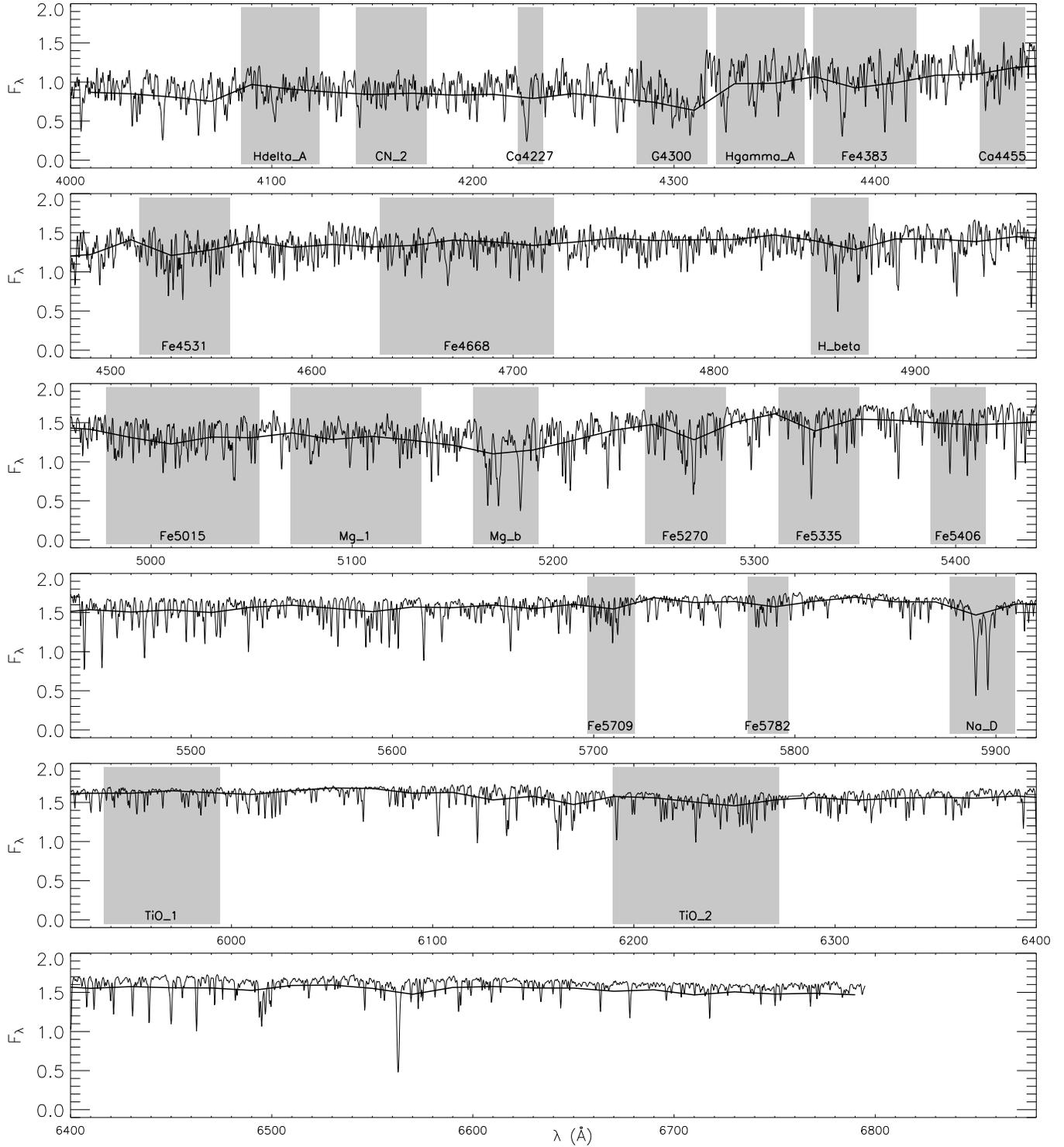}
    \end{center}
    \caption{\PEGASEHR{} high-resolution spectrum of a 10~Gyr-old  SSP of solar metallicity 
      compared to the low resolution spectrum from P\'EGASE.2. 
      The flux is given in units of
      $10^{-5}~\mathrm{L}_\odot$.\AA$^{-1}$.M$_\odot^{-1}$, where 
      $\mathrm{L}_\odot=3.826\times10^{33}$~erg.s$^{-1}$.
      Grey areas correspond to the passbands of some Lick indices.}
    \label{figure:spectrum}
  \end{figure*}
\end{center}

Figure~\ref{figure:spectrum} presents a prediction of a 10~Gyr-old
high resolution SSP spectrum at solar metallicity. 

The labeled grey areas in the plot identify the passbands of some Lick
indices \citep{Wortheyetal1994}.  We can notice that each passband
includes many lines of various chemical elements.  More precise line
identifications and their evolution are presented in the next section.
To highlight the spectral resolution improvement, we have overplotted
the same SSP computed with P\'EGASE.2 using the BaSeL library. 

It is worth noting that the adopted evolutionary tracks are
parametrized by their total metallicity $Z$, while the stellar library
is labeled by the iron abundance [Fe/H]. In the present work, we
assume the relation $\log_\mathrm{10}(Z/0.02)$ = [Fe/H].  
The  possible correlation  between metallicity
and relative abundances of $\alpha$-elements in the spectral library,
due to the selection of stars from the solar neighborhood, will be
invoked in the following to explain some of the differences with other models.

\begin{center}
  \begin{figure}[!tbf]
    \begin{center}
      \includegraphics{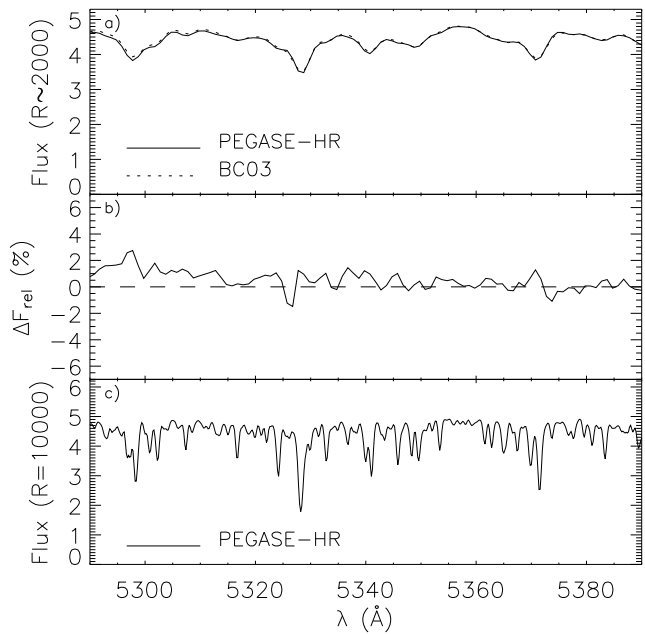}
    \end{center}
    \caption{Comparison of a portion of a smoothed \PEGASEHR{} SSP spectrum (\Zsol, 3~Gyr) with the model of \citet{BC03}. 
      The IMF is common to both models \citep{Salpeter}. 
      {\bf a)}~Both models overplotted, at the same low resolution $R\simeq 2\,000$. The flux is given in units of
              $10^{-5}$~L$_\odot$.\AA$^{-1}$.M$_\odot^{-1}$. 
      {\bf b)}~Relative difference of the fluxes at low resolution ($R\simeq 2\,000$)
      {\bf c)}~The original high resolution \PEGASEHR{} SSP spectrum ($R=10\,000$).}
    \label{figure:compbc}
  \end{figure}
\end{center}
%
%
\begin{center}
  \begin{figure}[!tbf]
    \begin{center}
      \includegraphics{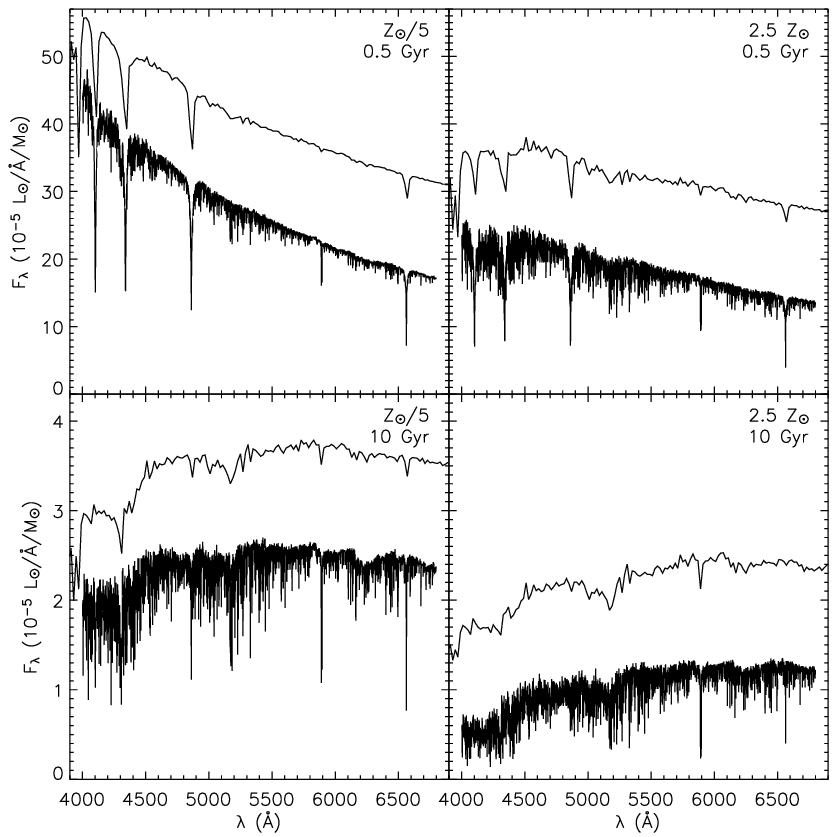}
    \end{center}
    \caption{Comparison of P\'EGASE.2 and \PEGASEHR{} spectra of SSPs, at metallicities $2.5\times$\Zsol{} and \Zsol/5, and ages 0.5~Gyr and 10~Gyr.
      The P\'EGASE.2 spectra are plotted with a vertical offset (a constant in each panel) for clarity.}
    \label{figure:individual_spectra}
  \end{figure}
\end{center}
%
%
\subsection{The spectra around specific stellar lines}
\label{section:elements}
\begin{center}
  \begin{figure*}[!tbf]
    \begin{center}
      \includegraphics{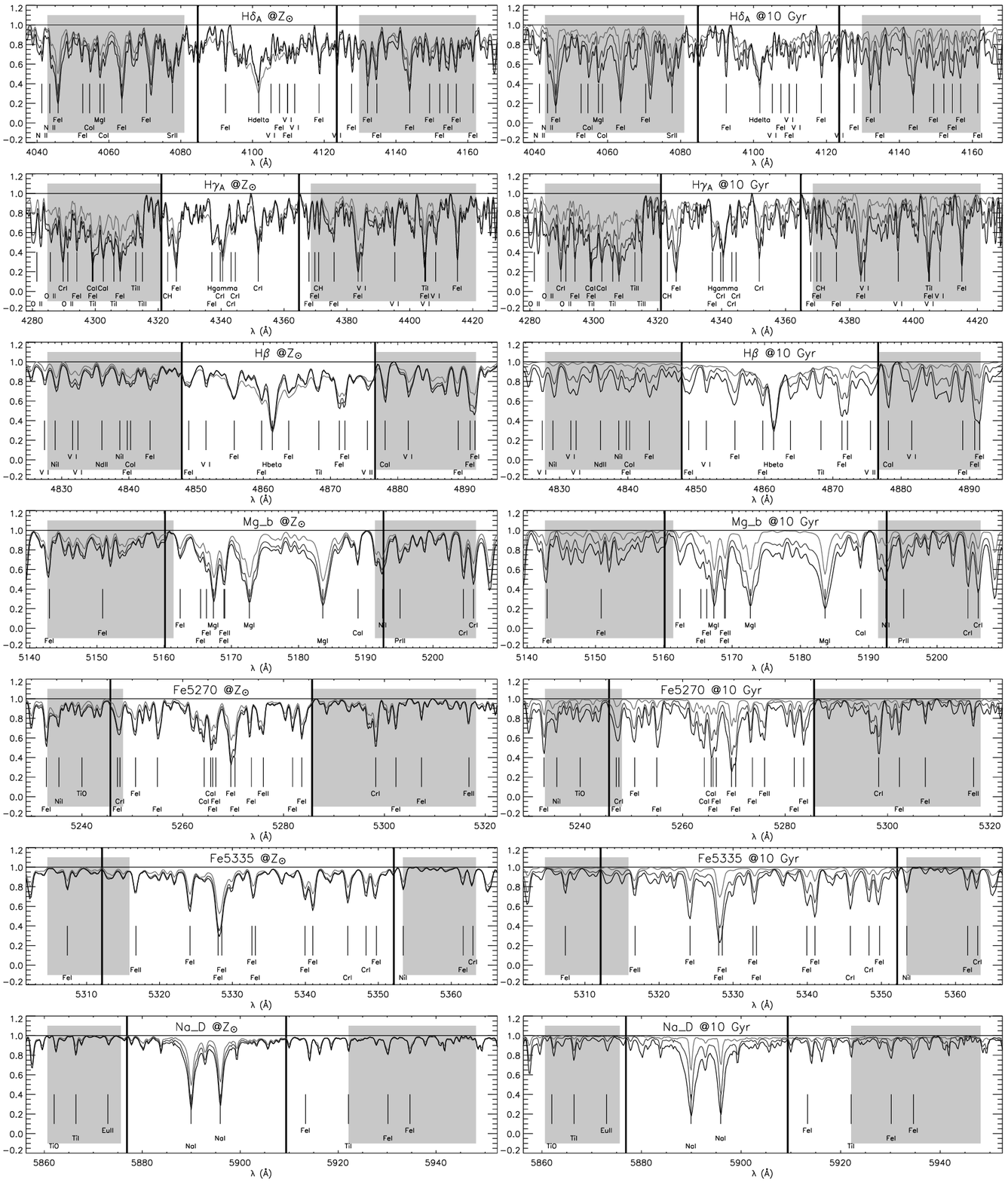}
    \end{center}
    \caption{Evolution of main stellar lines and continua for an SSP as a
      function of age (ages~=~1, 4, 13~Gyr, from light gray to black) at \Zsol{} (left)      
      and metallicity ([Fe/H]~= $-1.7$, $-0.4$, $0.4$, from gray to dark) at 10~Gyr
      (right).  Line identifications are from the revised ILLSS
      catalogue \citep{ILLSS}.  Grey areas show the blue and red
      pseudo-continua of the Lick indices and solid vertical
      lines delimit their central passband. In some
      cases (H$\gamma_A$, H$\delta_A$), significant evolution 
        with age is observed within the pseudo-continua (see text).}
    \label{figure:elements_superposed}
  \end{figure*}
\end{center}
%
%
Figure~\ref{figure:elements_superposed} presents the normalized
spectral distributions of SSPs centered on the main stellar lines
(H$\delta$, H$\gamma$, H$\beta$, Mg$_b$, Fe$\lambda$5270,
Fe$\lambda$5335, NaD).  The qualitative effects of age and
metallicity can be assessed: in the left column, all SSPs have solar
metallicity and their ages are at 1, 4, 13~Gyr; in the right column,
all SSPs are 10~Gyr-old and their metallicities are $\textrm{[Fe/H]}= -1.7$, $-0.4$,
$0.4$. Lick index passbands are shown with vertical lines and the grey
areas correspond to the pseudo-continuum passbands.  The widths of the Lick
passbands contrast with the narrow profiles of the stellar lines,
outlining the possibility of refined indices.

\subsection{The library of SSPs}
Further examples of synthetic  SSPs spectra at the resolutions 
of P\'EGASE.2 and P\'EGASE-HR are shown in
Fig.~\ref{figure:individual_spectra}. For clarity, 
P\'EGASE.2 spectra are offset vertically. The qualitative agreement in the
energy distribution and the broad features is very satisfactory.
A quantitative comparison is provided below.
Note that \PEGASEHR{}  and P\'EGASE.2 are based on the same algorithm, so that
the evolution of global parameters (metallicity, mass of gas, stellar mass and all
outputs independent of the library) is unchanged.

\section{Consistency of the \PEGASEHR{} outputs}
\label{section:previous}

\subsection{Quantitative comparison to P\'EGASE.2: luminosities and colors}
\label{section:delta}
\begin{center}
  \begin{figure}[!tbf]
    \begin{center}
      \includegraphics[width=8.5cm]{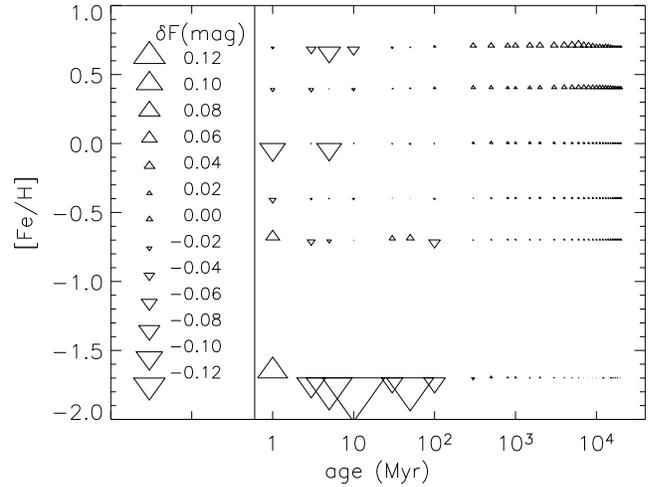}
    \end{center}
    \caption{Comparison of the integrated optical flux
       predicted by P\'EGASE.2 and \PEGASEHR{} for SSPs using $\delta F$ (4200-6700~\AA).
      The size of the symbols scales with magnitude difference 
      (see text in Sect.~\ref{section:delta} for details).}
    \label{figure:compFtot}
  \end{figure}
\end{center}

Figure~\ref{figure:compFtot} presents a comparison of the integrated
fluxes predicted by \PEGASEHR{} and by P\'EGASE.2 through the quantity
\[\delta F=-2.5 \times \log_\mathrm{10}
\left(
\frac{{\displaystyle
    \int _{\lambda=4200\mathrm{\AA}}^{6700\mathrm{\AA}} 
    F_\lambda^{\mathrm{PEGASE-HR}} \mathrm{d}\lambda}} 
     {{\displaystyle
	 \int _{\lambda=4200\mathrm{\AA}}^{6700\mathrm{\AA}}
    F_\lambda^{\mathrm{PEGASE.2}} \mathrm{d}\lambda}}
\right) , \]
for SSPs of various ages (1~Myr to 20~Gyr) and metallicities (from
$\textrm{[Fe/H]}=-1.7$ to $\textrm{[Fe/H]}=0.7$).  Absolute differences for ages greater than 10~Myr and [Fe/H]$>-1.7$ are always
smaller than 0.04 magnitudes.

\begin{center}
  \begin{figure}[!tbf]
    \begin{center}
    \includegraphics[width=8.5cm]{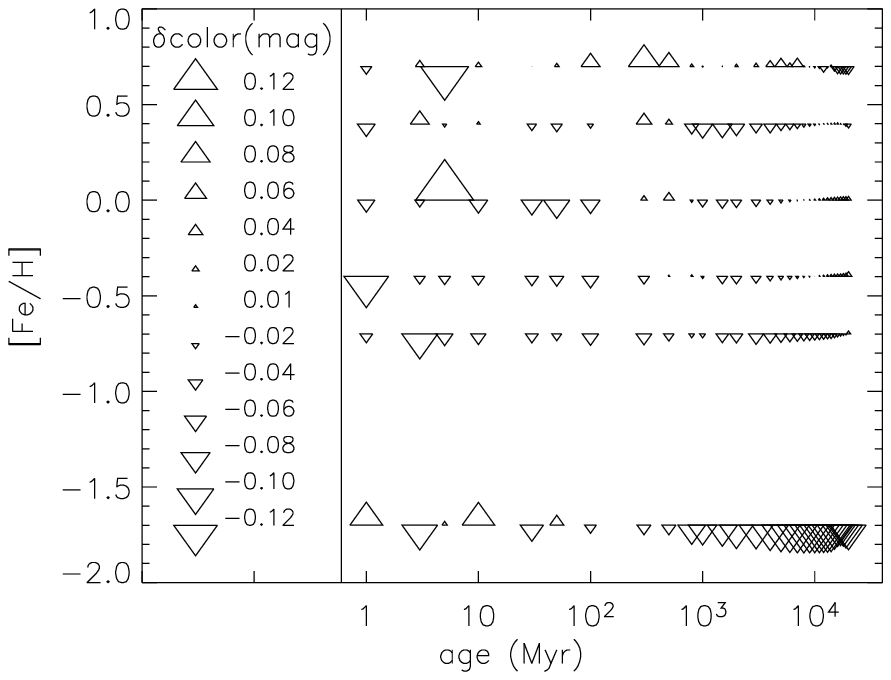}
    \end{center}
    \caption{Comparison of the optical colors
      predicted by \PEGASEHR{} and P\'EGASE.2 for SSPs.
      The adopted color compares the fluxes integrated over
      4300--4600\,\AA{} in the blue and 6400--6700\,\AA{}  in the red. The size of the
      triangles scales with the color difference expressed in magnitudes
      ($\delta{\textrm{color}}$) between the two models.
      Head-up triangles are plotted when \PEGASEHR{} produces
      a redder slope than P\'EGASE.2.}
    \label{figure:comp_coul_SSP}
  \end{figure}
\end{center}

Figure~\ref{figure:comp_coul_SSP} compares the slopes of the energy
distributions produced by the two versions of the code.  The slope is
measured with a color index, based on filters with rectangular
passbands spanning 4300--4600~\AA{} (``blue" band) and
6400--6700~\AA{} (``red" band).  Color differences between the two
models, over the whole range of metallicities and ages greater than 10~Myr, are lower than 0.12
magnitudes. Once more, the largest disagreements are confined to the youngest
ages and lowest metallicities.

Overall, the colors and fluxes of \PEGASEHR{} are consistent with those
of P\'EGASE.2 for ages greater than 10~Myr and [Fe/H]$>-1.7$. This
result confirms that the BaSeL and \'ELODIE libraries are very similar
at low resolution, at least over the range of parameters in which
stellar spectra contribute significantly to the optical light of a
galaxy. Such an agreement was not obvious \emph{a priori} since the
two libraries are constructed with very different methods: the former
is based on theoretical spectra, color-corrected to fit observed
stellar continua, whereas the latter is purely empirical.  As a
consequence, the low-resolution extended SEDs predicted by P\'EGASE.2
and the high-resolution spectra of \PEGASEHR{} can be used together to
refine SED studies.

%
\subsection{Comparison with Lick indices for SSPs}
\label{section:Lick}
Before studying lines at high resolution, we first want to check that 
our predictions at medium resolution are consistent with those of other models.

The traditional approach for the computation of Lick indices of SSPs
\citep[e.g][]{Wortheyetal1997,Bressanetal1996} is based on a library
of indices of individual stars and on analytic representations of this
set of data (the fitting functions). Thanks to the resolution of the
\'ELODIE spectra, our approach is more direct: we compute the
synthetic spectrum of a galaxy or SSP, degrade the spectral resolution
to the one of the Lick index definitions (8 to 11~\AA{} depending on
wavelength, as described by \citealp{Wortheyetal1997}), and then
measure the indices directly on the smoothed spectrum.  This allows us
to easily predict indices for any complex evolutionary scenario. In
the following, we compare our results for SSPs with those of other
groups.  

\subsubsection{Metal-sensitive indices}

\begin{center}
  \begin{figure*}[!tbf]
    \begin{center}

      \begin{tabular}{cc}
	  \includegraphics{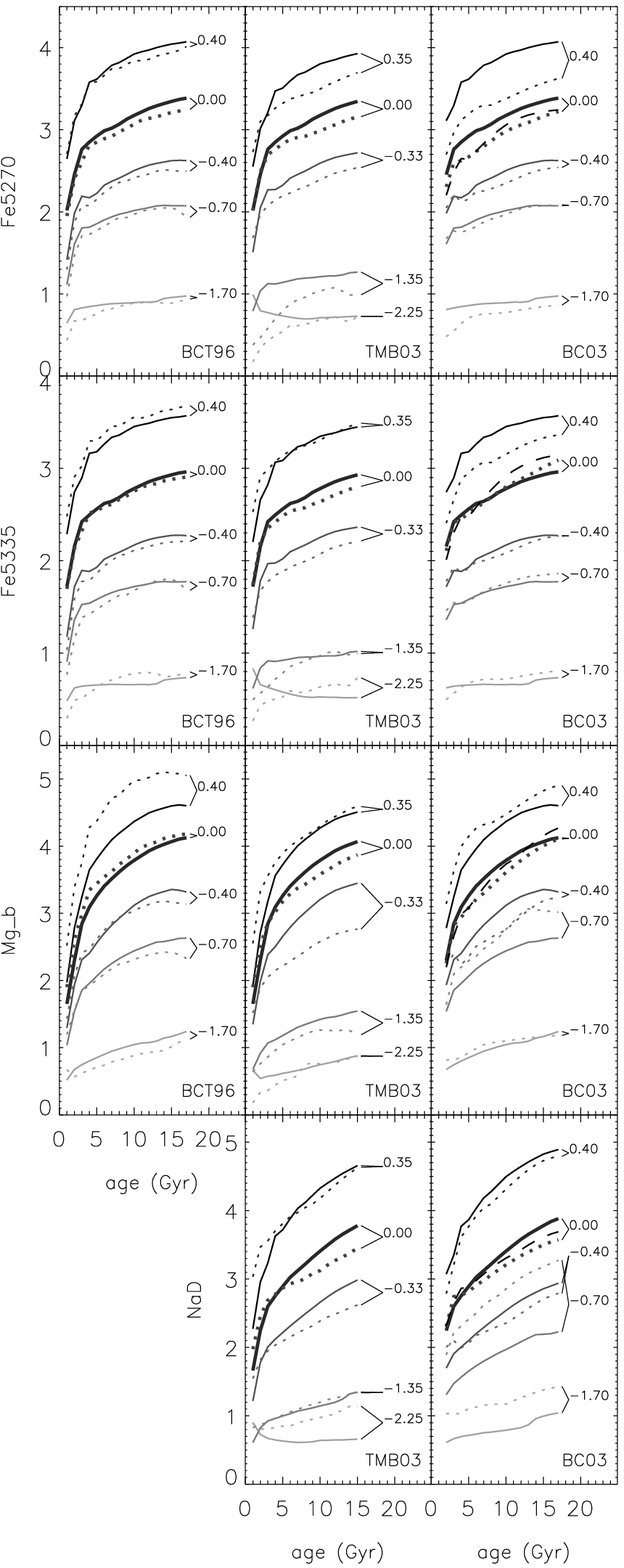}&
	  \includegraphics{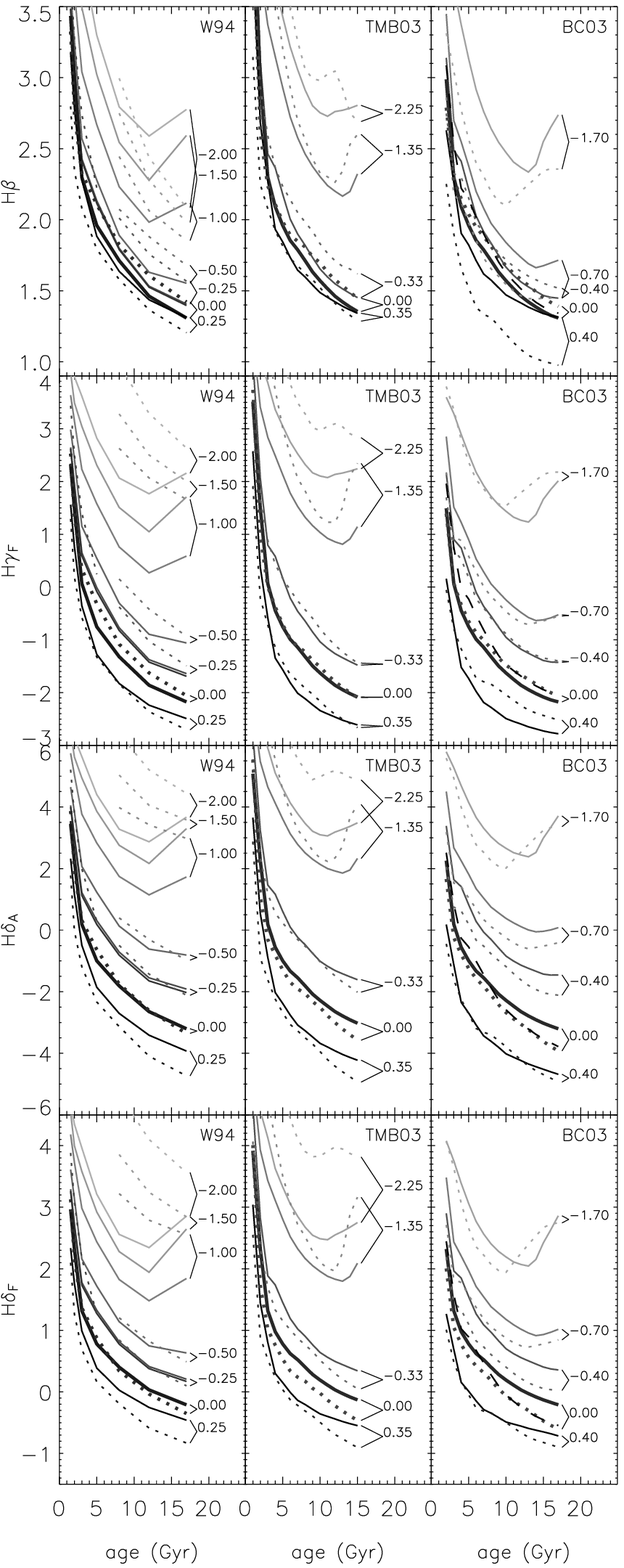}\\
      \end{tabular}

    \end{center}
    \caption{\PEGASEHR{} Lick indices (solid lines) for SSPs
      compared to models (dotted lines) of \citet{Wortheyetal1997} (W95), 
      \citet{Bressanetal1996}  (BCT96), \citet{Thomasetal2002} (TMB03) and \citet{BC03}
      (BC03). The figures on the right hand side of each panel
      represent the quantity [Fe/H].   For BC03 only, the
      dashed line shows the indices computed with the Geneva tracks
      instead of the Padova 1994 tracks, for solar metallicity.}
    \label{figure:fig_indices_met}
  \end{figure*}
\end{center}
The left hand side of Fig.~\ref{figure:fig_indices_met} shows the variations with
age of the metal-sensitive Lick indices Fe5270, Fe5335, Mg$_b$ and NaD, as computed
with \PEGASEHR, by \citet{Bressanetal1996}, by \citet{Thomasetal2002}, and by \citet{BC03}.
The comparisons are shown at various metallicities. 
\citet{Thomasetal2002} provide results for various relative abundances
of the $\alpha$-elements, and we have plotted those for $\textrm{[}\alpha\textrm{/Fe]}=0$.
The \citet{Salpeter} initial mass function (IMF) is used in the three models
(power law index $x=-1.35$). The lower and upper stellar masses
are adapted for each comparison.

\PEGASEHR{}, \citet{Bressanetal1996} and \citet{BC03} all use the
Padova stellar tracks in this comparison, while \citet{Thomasetal2002}
use what we will call the ``Cassisi tracks", a compilation from
\citet{Cassisietal1997,Bonoetal1997,Maraston1998,Salasnichetal2000}.
Moreover, both \citet{Thomasetal2002} and \citet{Bressanetal1996} 
use ``fitting functions'' to derive their synthetic indices, contrarily
to us and to \citet{BC03}.

The agreement of the iron indices between the three models  is quite good, 
except for the youngest SSPs
at the lowest metallicities. This shows that \'ELODIE spectra
and the iron index fitting functions of \citet{Wortheyetal1994}
are compatible over the range of parameters relevant for the 
synthesis of stellar populations, as expected from the 
direct comparison performed by \cite{ELODIE}. 
As stated in \citet{Maraston2002},
the Cassisi tracks should lead to slightly lower values of the
indices. This is what we observe, especially for iron indices. 
The discrepancy with \citet{BC03} at super-solar metallicity will be explained below.

The situation for the Mg$_b$ index is  more complex, since
magnesium is an $\alpha$-element. Both the \'ELODIE library and the
Lick fitting functions rely on an empirical stellar data set, in which
one should expect the anti-correlation between [Fe/H]
and [$\alpha$/Fe] characterizing the solar neighborhood
\citep{Fuhrmann1998,Fulbright2000}.  
\citet{Thomasetal2002} have included a correction for the
local anti-correlation between [Fe/H] and [$\alpha$/Fe] in their
sub-solar computations. At solar metallicity, where 
no such correction is required, their predictions and those of 
\PEGASEHR{} agree very well. As expected, \PEGASEHR{} produces a higher
Mg$_b$ index at sub-solar metallicities. We have observed that our predictions match better the
values given by \citet{Thomasetal2002} for [$\alpha$/Fe]=0.3.

The differences between the Mg$_b$ predictions of \PEGASEHR{} and those of
\citet{Bressanetal1996}, which are significant at super-solar
metallicities, can only result from selection effects in the
underlying stellar samples or in peculiarities of the interpolations
in regions of the HR diagram where only few stars are available.  
  Although the differences in stellar samples are probably the main
  source of disagreement between the models at sub-solar metallicity,
  the latter option is likely to explain the strong difference
  observed at $\textrm{[Fe/H]}=0.4$, as demonstrated below.

We have compared the Mg$_b$ indices measured on the \'ELODIE spectra
(after interpolation onto the grid of fundamental parameters used in
the population synthesis code) with those obtained for the same grid
of parameters from the fitting functions of \citet{Wortheyetal1994}.
The average offset is negligible, as can be expected from
\citet{ELODIE} and Fig.~\ref{figure:lick}.  However, the indices
measured on \'ELODIE spectra of non-giant stars (Mg$_b < 6$) have a
slight tendency to have lower values than those derived from fitting
functions for strong lines, and higher values for weak lines. This
effect, also present for the individual stars (see
Fig.~\ref{figure:lick}), is of the order of a few tenths of an
\AA{}ngst\"{o}m, i.e. not sufficient to explain the differences
observed for Mg$_b$ in Fig.~\ref{figure:fig_indices_met} between
\PEGASEHR{} and \citet{Bressanetal1996}.  More important is that the
Mg$_b$ fitting functions rise dramatically with decreasing
temperatures for low gravity red giant stars (\Teff$ < 3800$\,K).
When using the Padova stellar evolution tracks, these stars contribute
significantly to the optical emission of old metal-rich populations.
Very few such stars were present in the library used by
\citet{Wortheyetal1994}, all of them with sub-solar metallicity, and
their Mg$_b$ indices are high indeed (values above 15~\AA, up to
18.5~\AA).  The Mb$_b$ indices of the coolest super-solar giants in
the \'ELODIE library are lower by up to several \AA: the strongest
Mg$_b$ index for a giant is Mg$_b=14.3$~\AA{}. Also, there are only
two giant stars in the \'ELODIE library with [Fe/H]$>$0 and
Mg$_b>7$~\AA{} (HD18191 and HD169931). The interpolation of the
stellar spectra is very hazardous with so few stars in this range of
metallicity, effective temperature and surface gravity. Therefore, one
should not be surprised to observe strong differences in the indices
of $\alpha$-elements for old, super-solar stellar populations.

\begin{center}
  \begin{figure}[!tbf]
    \begin{center}
      \includegraphics{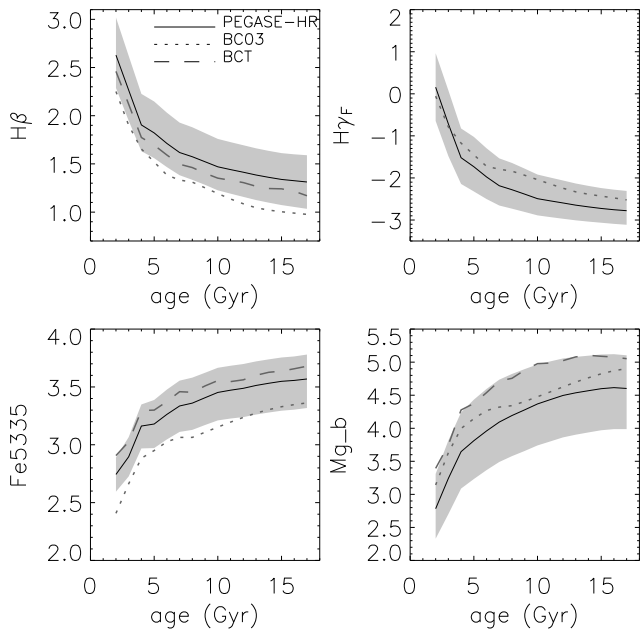}\\
    \end{center}
    \caption{Lick indices for $\textrm{[Fe/H]}=0.4$ SSPs. The indices of this
    work (solid line), \citet{BC03} (dotted line) and
    \citet{Bressanetal1996} (dashed line) are shown. The shaded region shows the
    uncertainty in the indices associated to the uncertainty in the
    effective temperature of the giant stars (see text for details).}
    \label{figure:fig_indices_teff}
  \end{figure}
\end{center}

Moreover, a comparison of our estimated effective temperatures with
the values published in \citet{Wortheyetal1994} for the stars in
common shows that the calibration of \Teff{} for the cool giant stars
is uncertain by about 150~K (5\%) which could also explain some of the
differences between the models at super-solar metallicities. The
effect of this uncertainty on the predicted indices at
$\textrm{[Fe/H]}=0.4$ is comparable to the discrepancy between the
models: Fig.~\ref{figure:fig_indices_teff} shows four indices of a
$\textrm{[Fe/H]}=0.4$ SSP obtained by shifting artificially the
temperatures of the stellar tracks by $\pm5$\% (shaded region). The
other models at this metallicity \citep{BC03,Bressanetal1996} predict
values compatible with this uncertainty.  However, this shaded region
probably overestimates the error due to the uncertainty in stellar
effective temperatures: the error in \Teff{} for the dwarf main
sequence stars is smaller than 5\%.  At solar or sub-solar
metallicities, the contribution of the giant stars to the spectrum of
an SSP is smaller. The uncertainty in \Teff{} then becomes negligible
in comparison to the uncertainty in the stellar tracks.

The index NaD is in quite good agreement with the other models, which
is remarkable because this index is contaminated by many telluric
lines. We took great care to remove these lines from the high
resolution stellar spectra. The non-monotonic evolution of the NaD
index with metallicity for old SSPs in the predictions of \citet{BC03} 
reflects the difficulty to model this index correctly.

\subsubsection{Hydrogenic indices, age tracers}

Balmer lines are generally considered to be excellent age tracers.
H$\alpha$ and H$\beta$ are often strong absorption lines, but they may
be filled in by circumstellar and interstellar emission. Thus, they provide
less reliable age diagnostics than H$\gamma$ and H$\delta$ which
are much less affected. 

To analyze stellar populations, \citet{Wortheyetal1997} defined Lick
indices around easily measurable Balmer lines: H$\beta_A$, H$\gamma_A$
and H$\delta_A$ for A stars with broadened wings, and H$\beta_F$,
H$\gamma_F$ and H$\delta_F$ for later types, with narrower Balmer
lines. They use the Revised Yale isochrones \citep{RYI} and
\citet{Vandenberg1985} isochrones extrapolated, when necessary, to
some parts of the HR diagram. The IMF adopted by them has a
\citet{Salpeter} slope, with $M_\mathrm{min}=0.1~M_\odot$ and
$M_\mathrm{max}=2~M_\odot$ (they model old populations only).
Figure~\ref{figure:fig_indices_met} (right hand side) shows an overall
satisfactory agreement between \PEGASEHR, \citet{Wortheyetal1997},
\citet{Thomasetal2002} and \cite{BC03} for metallicities
[Fe/H]$>$ -1.00.  However, we are also in rather good agreement with
\citet{BC03} for $\textrm{[Fe/H]}=-1.7$. We believe that some of the
discrepancies in the predictions of Lick indices for Balmer lines at
low metallicity are due to different treatments of the helium-burning
stars in the stellar tracks \citep{Charlotetal1996,Yi2003}.  This is
illustrated in the right panels of Fig.~\ref{figure:fig_indices_met}
where the \citet{BC03} predictions for solar metallicity are given
with the Padova tracks, but also with the Geneva
\citep{Geneve1,Geneve2} tracks. As for metal-sensitive indices, many of the discrepancies in
the predictions of the Balmer Lick indices, in particular at
super-solar metallicities, are likely due to differences in the stars
used to build the stellar libraries and in their estimated effective temperatures.

\section{Synthesis of evolved populations}
\subsection{Globular clusters}
Predictions of spectral feature strengths in SSP spectra can be directly
compared to observational data of coeval stellar populations, as found
in globular clusters. The data used here were obtained by
\citet{Bica1986a}, \citet{Huchraetal1996} and \citet{Trageretal1998}.
\paragraph{Equivalent width of the H$\beta$ line\\}
Equivalent widths (EW) of the H$\beta$ line were measured as a function of age
and metallicity for a sample of 
63 observed star clusters in the Galaxy, the LMC, and the
SMC by \citet{Bica1986a, Bica1986b}. 
We predict the evolution of the H$\beta$ EW with \PEGASEHR,
with the same spectral resolution (12~\AA, Gaussian smoothing) and 
the same EW definition:
 \begin{equation}
  EW=\int_W \left[ 1- F(\lambda)/F_c(\lambda) \right] \textrm{d}\lambda
\end{equation}
where $F_c$ is the continuum in the middle of the window $W$.  The
continuum is the upper envelope of the spectrum: maximal fluxes around
the points $\lambda\lambda=4570,~5340,~6630$~\AA~($\pm 20$~\AA) are
connected by straight lines. EW is given in \AA.

\begin{center}
  \begin{figure}[!tbf]
    \begin{center}
      \includegraphics[width=8.8cm]{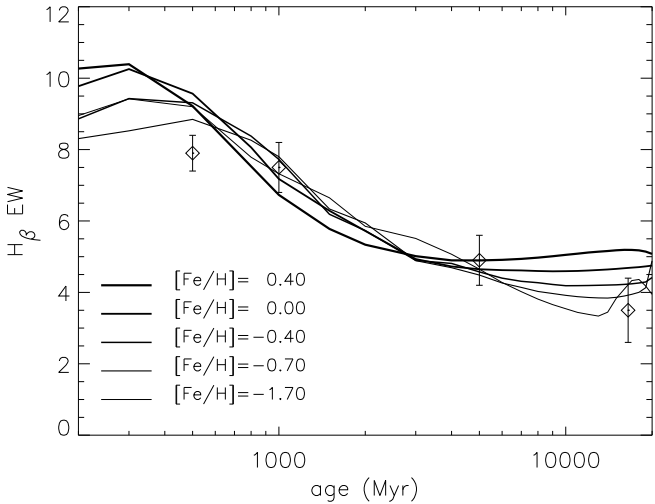}
    \end{center}
    \caption{Evolution sequences with age of the H$\beta$ equivalent width from \PEGASEHR{}  for various 
      metallicities (solid lines) compared to the binned globular
      cluster data \citep{Bica1986a,Bica1986b}.}
    \label{figure:Hbeta}
  \end{figure}
\end{center}

Figure~\ref{figure:Hbeta} plots the binned data compared to the \PEGASEHR{} 
sequences of H$\beta$ equivalent widths for various
metallicities ($\textrm{[Fe/H]}=-1.7$, $-0.7$, $-0.4$, $0.$, $0.4$). Note that the
isochrones used by \citet{Bica1986b} to estimate ages differ from
the ones used here, allowing no very precise age comparison.

The models agree with the data at ages roughly between 1 and 10~Gyr.
At younger ages, emission lines might partly fill in absorption lines in the 
observations, and ongoing star formation may rule out the model of 
the strictly instantaneous ($<$1~Myr) burst described by an SSP.
At the oldest ages, the metal deficiency of globular clusters is 
extreme, and we reach the regime in which our interpolated spectral
library suffers from larger uncertainties.

\begin{center}
  \begin{figure}[!tbf]
    \begin{center}
      \includegraphics[width=8.5cm]{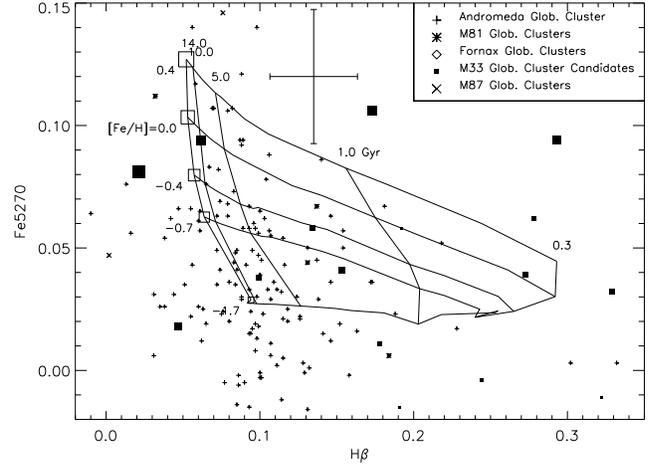}
    \end{center}
    \caption{Fe5270 vs. H$\beta$ indices for 193~extragalactic globular clusters \citep{Huchraetal1996}. 
      Solid lines are \PEGASEHR{} SSP models for several ages and
      metallicities, as labeled on the sequences. The typical error
      bar for the data is plotted above the grid. For M33 globular
      clusters, the size of the full squares indicates [Fe/H]
      \citep[as determined by][]{Huchraetal1996}, following the [Fe/H]
      scale given by the empty squares plotted on the 14~Gyr
      isochrone.}
    \label{figure:Huchra_1}
  \end{figure}
\end{center}

\paragraph{The iron-H$\beta$ diagram\\}
Among the 13 indices measured in 193 extragalactic globular clusters
by \citet{Huchraetal1996}, we select the index Fe5270 as a
representative metallicity tracer and H$\beta$ as the best available
age tracer (see also Table \ref{table:deltaEW}). The definitions of
indices in \citet{Huchraetal1996} differ from the Lick ones:
\begin{equation}
  I=-2.5 \log \left[ \frac{2 F_I}{(F_{C1}+F_{C2})} \right]
\end{equation}
where $F_I$, $F_{C1}$ and $F_{C2}$ are mean fluxes in the passband,
blue and red continua respectively. The passbands are also different
from the Lick ones. In Fig.~\ref{figure:Huchra_1}, the grid of
\PEGASEHR{} models (with ages=0.3--14~Gyr and $\textrm{[Fe/H]}=-1.7$ to
0.4) is compared to the data sample. The grid encompasses a wide range
of data, when taking into account the large observational error bar.
However, several points (mostly M31 clusters) are outside the
theoretical grid. Assuming that the behavior of the indices of
\citet{Huchraetal1996} is comparable to that of the corresponding Lick
indices, the offset in H$\beta$ for the lowest metallicity clusters is
consistent with the uncertainties already mentioned for this regime.
For M33 clusters,
\citet{Brodieetal1991} give estimations of the individual
metallicities.  Overall, we find a rather good agreement between 
measured metallicities of clusters and the grid. This
agreement is illustrated by the size of the squares in the plot
(filled squares for M33 data and empty squares for the model
predictions).

\begin{center}
  \begin{figure}[!tbf]
    \begin{center}
      \includegraphics[width=8.5cm]{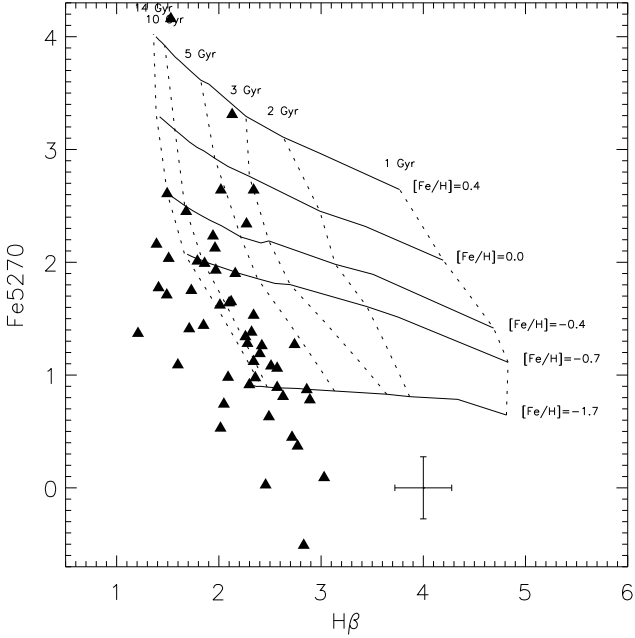}
    \end{center}
    \caption{Fe5270 vs. H$\beta$ indices for 50 globular clusters of 
      the Milky Way and M31 \citep{Trageretal1998}.
      The grid corresponds to \PEGASEHR{} SSP models, with iso-age
      sequences (dotted lines) and iso-[Fe/H] sequences (solid lines).
      The median error bar for the data is plotted below the grid.}
    \label{figure:Trager}
  \end{figure}
\end{center}

More recently, some precise measurements of Lick indices were made on
globular clusters of the Milky way and M31 and were compiled by \citet{Trageretal1998}. 
The measurement error is smaller for these
data, and enables a more precise determination of
ages. Figure~\ref{figure:Trager} shows the observed values of Fe5270
and H$\beta$ together with our synthetic grid for these indices. It appears that
the locus of the globular clusters in the grid is
approximately the same as for the extragalactic clusters of Fig.~\ref{figure:Huchra_1}. 
However, the ages are
determined much more precisely and they span a narrower range (3--20~Gyr) with
mainly sub-solar metallicities.

\subsection{Elliptical galaxies}
Elliptical galaxies are often modeled as purely coeval stellar
populations since the bulk of their stellar population formed within a
very short timescale (less than 1~Gyr).  We hereafter prefer to use the
typical scenarios of the model P\'EGASE \citep{FRV97} including infall
and galactic winds. The star formation rate is proportional to the mass of gas, with
respective timescales of 1~Myr for starburst, less than 1~Gyr for
elliptical and up to 10~Gyr for spiral galaxies. 

\begin{center}
  \begin{figure}[!tbf]
    \begin{center}
    \includegraphics[height=20cm]{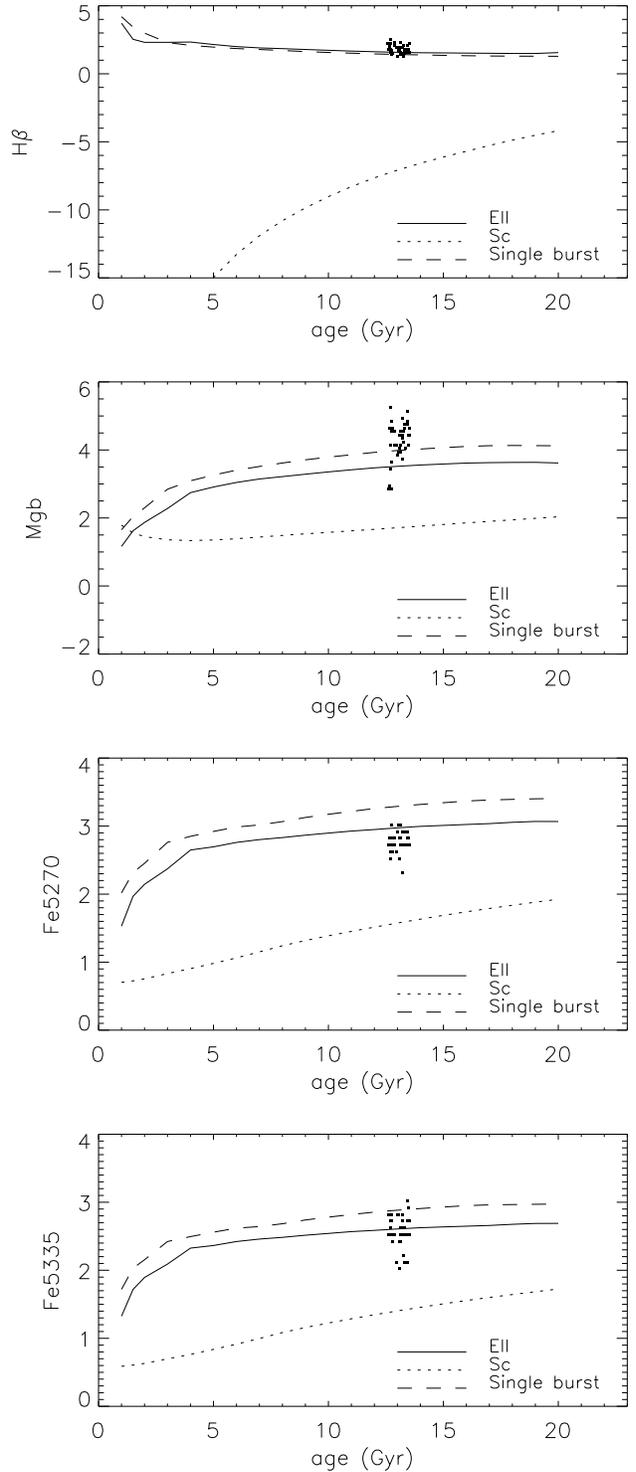}
    \end{center}
    \caption{Lick indices of \PEGASEHR{} models for elliptical, Sc and
      single burst galaxies together with data from local elliptical
      galaxies \citep{Trager}.  The age of the data is arbitrarily set by us to
      13~$\pm$~0.5~Gyr, randomly distributed (for the clarity of the
      plot). The presence of emission lines associated to ongoing star
      formation explains the low values of the H$\beta$ index
      predicted for evolved spiral galaxies.}
    \label{figure:indices_ell}
  \end{figure}
\end{center}

Figure~\ref{figure:indices_ell} shows the predictions of the main Lick
indices with our three scenarios \citep[spiral, elliptical, and instantaneous burst, 
as defined in][]{ZPEG} compared to the same indices measured
on local elliptical galaxies \citep{Trager}. The assumed galaxy age is
13$\pm$0.5~Gyr, but the indices are not very sensitive to age in this
range.  Mg$_b$ predictions are systematically low, which
is consistent with the known over-abundance of $\alpha$-elements in
elliptical galaxies.

\section{A systematic search for new narrow indices}
\label{section:newindices}
Lick indices are considered, to a certain extent, to be good estimators
for age (H$\beta$ index) or metallicity (Fe or Mg$_b$ indices).  Yet
these indices are defined for medium resolution (FWHM $\sim 8$~\AA):
many absorption lines, from various elements, are included in each
passband. The full resolution power of \PEGASEHR{} makes it possible to build new
narrow indices for a better estimation of ages and metallicities. At
the limit, indices could be as narrow as a few tenths of an \AA{}ngstr\"{o}m,
but they will be accessible only to a few instruments. Also, the
exposure time would have to be very high to obtain a high S/N ratio.
Moreover, stellar velocity dispersion broadens the lines (50
km\,s$^{-1}$ corresponding to FWHM=2.0~\AA{} at 5000 \AA{} for example).
Extremely narrow indices would turn out to be unusable.  For these
reasons, we restrain our search to 2~\AA-wide indices. In order to
disentangle age and metallicity, we tried to build couples of indices
sensitive to either age or metallicity, producing the most
orthogonal and regular grids in the age-$Z$ plane, with the smallest
measurement errors at a fixed mean S/N ratio (hereafter 50). 
Indeed, as emphasized by \citet{Cardieletal2003}, this 
association of small measurement errors (which depend directly 
on the definitions of the indices)  with an orthogonal grid is 
the most promising combination for disentangling age and metallicity signatures.
The errors in index measurements are estimated following the formulae of 
\citet{Cardiel1998}.

\subsection{Evolution of the pseudo-continuum}
\label{section:continuum}

\begin{center}
  \begin{figure}[!tbf]
    \begin{center}
    \includegraphics[width=8.5cm]{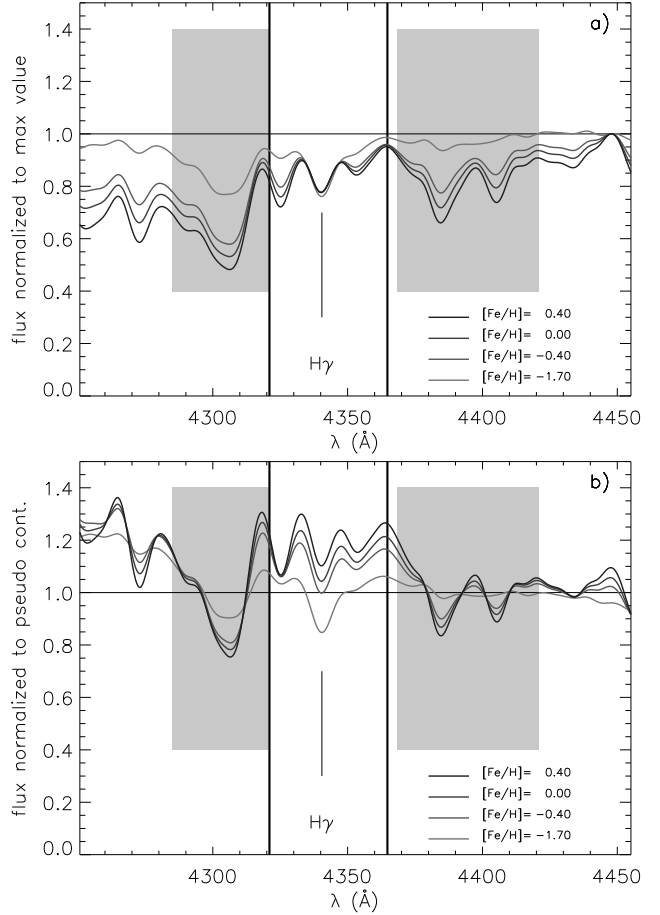}
    \end{center}
    \caption{Evolution of the H${\gamma}$ line as a function of metallicity for a 10~Gyr-old starburst:
      {\bf a)} normalization at 4448\AA; {\bf b)} normalization to the
      Lick H$\gamma_A$ pseudo-continuum.  In both panels, the
      spectrum is degraded to the Lick resolution. Grey zones
      represent the H$\gamma_A$ Lick blue and red continuum
      definitions and solid vertical lines delimit the H$\gamma_A$
      feature passband.}
    \label{figure:continus}
  \end{figure}
\end{center}

The definition of the continuum is an important aspect of any index
definition. The continuum can be defined, as in \citet{Bica1986a}, by the upper
envelope of the spectrum, but is then strongly sensitive to
noise and emission lines. The Lick indices define a``pseudo-continuum'': 
a straight line between the mean fluxes in a blue and a red window. 
This choice is more robust and is convenient for the analysis of
observations.  
The Lick normalization is empirical because the pseudo-continuum may evolve 
by itself with age and metallicity of the underlying populations.
Fig.~\ref{figure:continus} shows this effect on the H$\gamma$ line.
SSP spectra at various metallicities are presented at 10~Gyr with the Lick
index resolution. Two normalizations are presented: 
\begin{enumerate}
\item[a)]  at 4448~\AA{} (top panel);
the blue part of the Lick pseudo-continuum dramatically changes
because it falls in the G band;
\item[b)] at the Lick pseudo-continuum (bottom panel); the variations seem 
to reside in the H$\gamma$
line itself.
\end{enumerate}
 Depending on normalization, conclusions may strongly differ.
%

\subsection{A systematic search method}
We have developed a method to find narrow indices, defined like the
Lick indices, that could be good estimators of age or of metallicity.
The first step is to compute SSP spectra for various metallicities,
and various ages. Then, we normalize each spectrum to the maximum
value within a sliding 50~\AA-wide window. We thus obtain spectra
with almost flat continua. Next, we measure, for each wavelength, the
relative flux variations induced by either age changes or metallicity
changes.  We use the relative differences between these variations to
select narrow feature windows and pseudo-continuum windows; we thereby
define indices particularly sensitive to age or to metallicity.  Since
the S/N ratio of the stellar library is finite, we take into account
the uncertainties in the measurement of the flux variations relative
to age or to metallicity. This leads us to select easily measurable
indices only (i.e. with small measurement error bars).
\subsection{The age-sensitive H$\gamma$ index H$\gamma$\_VHR  }
\label{HgVHR}
The H$\gamma$ line, as a Balmer line, is a good age-sensitivity
tracer. Moreover, unlike H$\alpha$ and H$\beta$, it is seldom filled
in by nebular emission (H$\gamma/$H$\beta\simeq0.5$ in \ion{H}{ii}
regions).  The upper panel of Table \ref{table:newindices} presents
various definitions for a H$\gamma$ index (wavelengths and
resolutions): H$\gamma$\_F (Worthey, Lick resolution $\simeq$ 8~\AA),
H$\gamma$\_Vaz (\citet{Vazdekis1999}, $\sigma=200$~km\,s$^{-1}$: FWHM
$\simeq 6.8$~\AA), H$\gamma$\_HR (\citet{Jones1995},
$\sigma=80$~km\,s$^{-1}$: FWHM $\simeq 2.7$~\AA).  We hereafter propose
the new index H$\gamma$\_VHR (VHR standing for Very High Resolution)
with FWHM=2.0~\AA{} ($\sigma\simeq$60 km\,s$^{-1}$), well suited to \PEGASEHR.  

\begin{center}
  \begin{figure}[!tbf]
    \begin{center}
    \includegraphics[height=19.0cm]{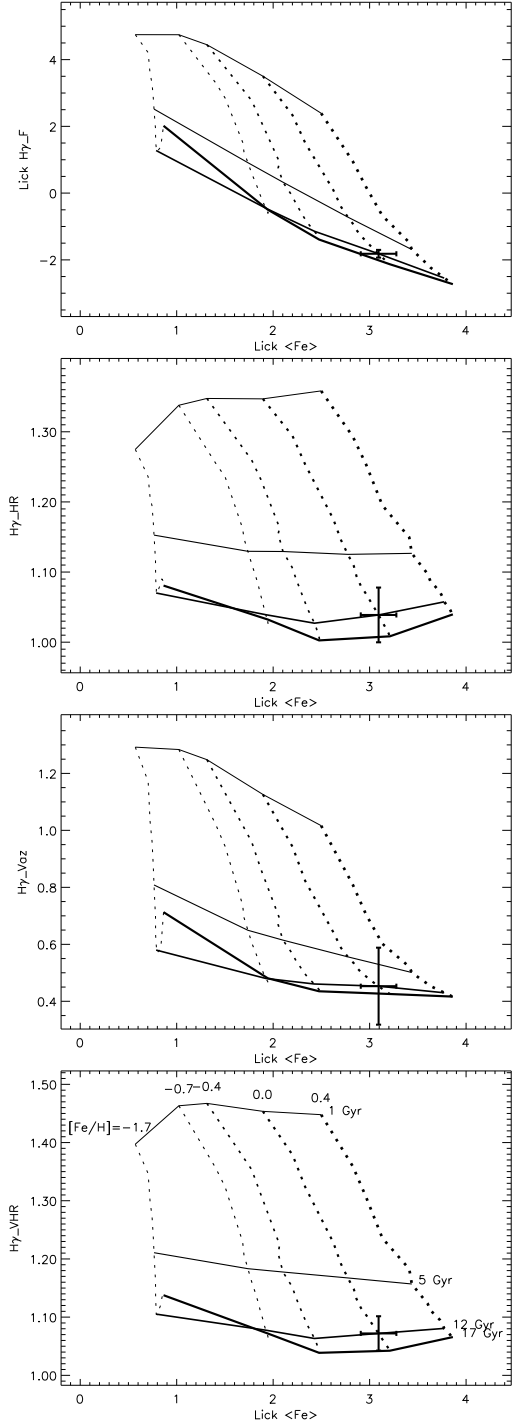}
    \end{center}
    \caption{H$\gamma$ indices vs. $<$Fe$>$ Lick index. From top to
      bottom, the H$\gamma$ indices are computed with the definitions
      of Worthey, Jones, Vazdekis, and our Very High Resolution index
      (see text for details). Error bars are computed from
      \citet{Cardiel1998} with an average S/N ratio=50. Solid lines
      are iso-age sequences (1, 5, 12, 17~Gyr). Dotted lines are
      iso-[Fe/H] sequences ($-$1.7, $-$0.7, $-$0.4, 0., 0.4). 
      The thickness of the lines increases with age and [Fe/H]. Notice the
      particularly flat iso-age sequences of the H$\gamma$\_VHR index, which
      indicates a very low $Z$-sensitivity.}
    \label{figure:fig_Balmer_indices}
  \end{figure}
\end{center}

Figure~\ref{figure:fig_Balmer_indices} presents the comparison of the
age-$Z$ grids built up with various H$\gamma$ indices for the
\emph{age} axis, and with the Lick $<$Fe$>$ index for the
\emph{[Fe/H]} axis.  The last grid (H$\gamma$\_VHR vs.  $<$Fe$>$)
shows that the H$\gamma$\_VHR index is particularly insensitive to the
metallicity and allows a slightly better age estimation than H$\gamma$\_HR at
low metallicity. For old, solar-metallicity stellar populations, it is very similar to 
the H$\gamma$\_HR index, but the error bar associated to its measurement (at S/N=50 on the plot) is smaller.
It may be measured at high ($>50$) S/N ratio with the new generation of instruments.
\begin{table*}[!tbf]
\centering
\caption[]{Various H$\gamma$ indices and resolutions (last column). 
H$\gamma$\_VHR and H$\gamma$\_Z are the new ones that we propose (see Sect.~\ref{HgVHR} and \ref{HgZ}).}
\begin{tabular}{lcccc}
\hline
\hline
Name             &    Feature           &         Blue band
                 &    Red band          &    FWHM (\AA) \\
\hline
H$\gamma$\_F      & 4332.500 4353.500     &  4284.750 4321.000    &
                 4356.000 4386.000      &      9.5\\
H$\gamma$\_HR     & 4338.607 4342.347     &  4333.000 4335.000    &
                 4348.000  4350.000      &       2.7\\
H$\gamma$\_Vaz    & 4332.000 4352.250     &  4331.000 4340.750    &
                 4359.250 4368.750        &     6.8\\
\hline
H$\gamma$\_VHR    &   4337.600     4341.200  &   4327.600     4336.400   &
              4348.800     4366.800    &    2.0\\
H$\gamma$\_Z      & 4351.000 4354.000     &  4318.000 4322.500    &
                 4362.000 4365.000        &    2.0\\
\hline
\end{tabular}
\label{table:newindices}
\end{table*}
%
%
%
\subsection{A metal-sensitive H$\gamma$ index: H$\gamma$\_Z}
\label{HgZ}
\begin{center}
  \begin{figure}[!tbf]
    \begin{center}
    \includegraphics{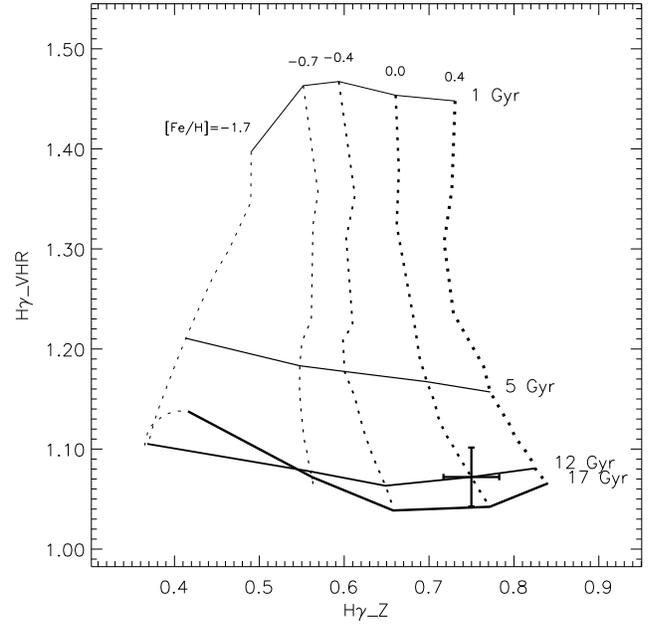}
    \end{center}
    \caption{H$\gamma$\_VHR index (age tracer) vs. H$\gamma$\_Z index
      (metallicity tracer) indices. Error bars are computed from
      \citet{Cardiel1998} with an average S/N ratio of 50. Same symbols as
      in Fig.~\ref{figure:fig_Balmer_indices}. }
    \label{figure:fig_HH}
  \end{figure}
\end{center}

The high spectral resolution (FWHM=2.0~\AA) makes it possible to present the new
H$\gamma$\_Z index, highly sensitive to metallicity. This might look
surprising, but \citet{Vazdekis1999} and \citet{Jones1995} already
noted that many metallic lines are present in every Lick index.  We
call this new index H$\gamma$\_Z because it is very close to the
H$\gamma$ line, but the central passband actually measures a Chromium
(Cr) line. This index is voluntarily defined for a spectral
resolution corresponding to FWHM=2.0~\AA. This narrow index is quite
complementary to the index H$\gamma$\_VHR: the passbands are very close to
each other, and a single short spectrum is enough to perform an age-Z
diagnostic. The independence of the parameters traced by the two
indices is exemplified in Fig.~\ref{figure:fig_HH}. The diagram
H$\gamma$\_VHR (age-sensitive) vs. H$\gamma$\_Z ($Z$-sensitive) should
be useful to interpret high-resolution spectra.\\
\subsection{$Z$ and age sensitivities}
We present in Table \ref{table:deltaEW} updated values of the
sensitivities to age and metallicity of some Lick indices
(H$\gamma$, H$\beta$, Fe5335, Fe5270) and of the old and new
H$\gamma$ indices.  We use the definition of \citet{Worthey1994} for
the quantity $\left( \Delta_\mathrm{age}/\Delta_\mathrm{[Fe/H]}
\right)$ called ``$Z$-sensitivity''. It is the ratio of the fractional
change in age to the fractional change in $Z$ needed to induce the
same variation of the index value.  A value close to zero means that
the index is completely insensitive to metallicity variations, whereas
a large absolute value indicates that the index is insensitive to age
variations.  We also give these fractional changes
$\Delta_\mathrm{age}$ and $\Delta_\mathrm{[Fe/H]}$ for a 10\%
variation of the index value around the zero point (\Zsol,~12~Gyr).  A
comparison to values obtained with previous models
\citep{Worthey1994,Jones1995} is also given in the last column.  Our
new index H$\gamma$\_VHR is quite insensitive to metallicity
variations around the zero point. 
The new index H$\gamma$\_Z is very sensitive to
variations of metallicity, even more than the Mg$_b$ index.
%
\begin{table}[!tbf]
\centering
\caption[]{Sensitivity of line equivalent widths to age (1--13~Gyr) and
  metallicity ($\textrm{[Fe/H]}=-1.7$ to $0.4$). $\Delta_\mathrm{[Fe/H]}$ represents the
  fractional change in [Fe/H] needed to produce a 10 \% variation of
  the index value at the zero point (\Zsol,~12~Gyr).
  $\Delta_\mathrm{age}$ is the similar quantity computed with the age.
  $\frac{\Delta_\mathrm{age}}{\Delta_\mathrm{[Fe/H]}}$ is the ratio of fractional
  change in age to the fractional change in [Fe/H] required to produce the
  same change in the index value \citep{Worthey1994,Jones1995}.
  High absolute values correspond to high sensitivity to $Z$.  This quantity is
  evaluated around the (\Zsol,~12~Gyr) zero point.  The last column
  reproduces the values in Table 6 of \citet{Worthey1994} and in Table~2
  of \citet{Jones1995}.}
\begin{tabular}{lcccc}
\hline 
\hline 
Index & 
$\Delta_\mathrm{[Fe/H]}$ & $\Delta_\mathrm{age}$ & $\frac{\Delta_\mathrm{age}}{\Delta_\mathrm{[Fe/H]}}$ & 
$\left( \frac{\Delta_\mathrm{age}}{\Delta_\mathrm{[Fe/H]}} \right)_{\textrm{\tiny{W,J}}}$\\ 
\hline
Fe5335                  & 0.41 & 0.75 & 1.8 & 2.8\\  
Fe5270              	& 0.42 & 0.74 & 1.8 & 2.3\\  
{\bf H$\gamma$\_Z}  	& 0.73 & 1.03 & 1.4 & \\     
Mg$_b$             	& 0.55 & 0.56 & 1.0 & 1.7\\  
H$\delta_A$         	& 0.16 & 0.14 & 0.9 & 1.1\\  
H$\gamma_A$         	& 0.36 & 0.31 & 0.9 & 1.0\\  
H$\gamma_F$         	& 0.23 & 0.15 & 0.7 & \\     
H$\gamma$\_Vaz      	& -3.04 & -0.42 & 0.1 & \\   
H$\beta$            	& -1.62 & -0.26 & 0.2 & 0.6\\
H$\gamma$\_HR       	& 6.50 & -1.06 & -0.2 & 0.0\\
{\bf H$\gamma$\_VHR}	& 11.05 & -1.06 & -0.1 & \\  
\hline
\end{tabular}
\label{table:deltaEW}
\end{table}
%
%
\section{Discussion and Conclusion}
\label{section:conclusion}
The code \PEGASEHR{} is a spectro-photometric evolution model of
galaxies with very high spectral resolution in the optical. It
inherits the P\'EGASE.2 accurate modeling of galaxy evolution, as
demonstrated by the compatibility of colors and low-resolution spectra
between the two models. The \'ELODIE library is complete enough to give an extensive coverage
of the HR diagram. Efforts are currently made to complete the
library with low metallicity stars.  Its quality depends on the
estimation of stellar parameters with TGMET and on the interpolator
which will be soon published. The \'ELODIE archive continues to grow
from new observations, and the density of stars in the parameter space
(\Teff, $\log_\mathrm{10} g$, [Fe/H]) increases. The improvement will
be particularly noticeable for fast evolving stars and a better
accuracy will be obtained for young starbursts (age~$<$~10~Myr).

We checked that \PEGASEHR, degraded at low resolution, is in good
agreement with other recent models: P\'EGASE.2,
\citet{Bressanetal1996}, \citet{Thomasetal2002}, or \citet{BC03}. The predicted
evolution of Lick indices for Balmer lines and metallic lines is
compatible with most other models and observations. 
The small number of cool giant stars with super-solar metallicity 
in the \'ELODIE library and the uncertainty on their effective temperature 
makes the predictions of indices quite uncertain 
for old, metal-rich stellar populations.
The influence of the non-solar enrichment for $\alpha$-elements in the
\'ELODIE library needs more investigation.

We systematically explored the wavelength domain to find new high
resolution indices, sensitive to either age or metallicity.  Inside
the passband of the classical H$\gamma$ Lick index we find two very
high resolution indices: one is sensitive to age only, and is quite
similar to the index of \citet{Jones1995}. The other one is sensitive
to metallicity only.  These 2~\AA-wide features are defined for
velocity dispersions up to 60~km\,s$^{-1}$. We note that these indices,
despite a systematic investigation at high resolution, are not
fundamentally different from the previously existing ones. This may
indicate that future improvement will mainly come from high-resolution
SED fitting rather than from classical indices.

This model is suited to the analysis of high resolution spectroscopic
observations with rest-frame wavelengths falling in the
4000--6800~\AA{} interval. In particular, the most recent spectrographs
on large telescopes (GIRAFFE, VIMOS or ISAAC at the VLT, GMOS or GNIRS
at the Gemini Observatory, EMACS on Magellan, TWIN at Calar Alto, ISIS at WHT\ldots) will
benefit from this model. We already presented tests of inversion
methods showing their ability to separate disk and bulge
components from a spectrum, deriving simultaneously their population
characteristics and their kinematics
\citep{Prugnieletal2003,Ocvirketal2003}.  This approach will be
fundamental in many scientific applications, such as the study of
bulges for which correcting the disc contamination is
a pre-requisite \citep{Prugnieletal2001}, or of high surface brightness dwarf objects.

The code, as well as SSPs, can be downloaded on
the P\'EGASE web site at \texttt{http://www.iap.fr/pegase/}. Because
the output spectra now contain ten times as many wavelength points as
the P\'EGASE.2 spectra, the SSPs are provided in FITS format. They contain
all the information previously given by P\'EGASE.2 outputs, except the colors. 
Measurements of Lick indices are included. A code to convert the FITS files
into the P\'EGASE.2 {\sc ASCII} format is also available for the user's
convenience.

We stress that this new tool is unique for studying stellar populations in
nearby and distant galaxies observed with the new high resolution
spectrographs.  Other improvements will come from 
multi-wavelength analyses, by
combining \PEGASEHR{} optical spectra with far-ultraviolet and
near-infrared SEDs. Another improvement will come from the implementation of 
non-solar abundances in the evolution. For this purpose, we are
presently measuring chemical abundances from the high resolution
spectra of the \'ELODIE library.

\begin{acknowledgements}
We would like to thank the referee, Guy Worthey, for his very useful comments which helped improve the paper significantly.
This work was supported in part by the French Programme National Galaxies.
\end{acknowledgements}

\end{document}